# Evaluation of ARM CPUs for IceCube available through Google Kubernetes Engine

*Igor* Sfiligoi[1], *David* Schultz[2], *Benedikt* Riedel[2] and *Frank* Würthwein[1]

[1]University of California San Diego, La Jolla, CA 92093, USA
[2]University of Wisconsin–Madison, Madison, WI 53715, USA

**Abstract.** The IceCube experiment has substantial simulation needs and is in continuous search for the most cost-effective ways to satisfy them. The most CPU-intensive part relies on CORSIKA, a cosmic ray air shower simulation. Historically, IceCube relied exclusively on x86-based CPUs, like Intel Xeon and AMD EPYC, but recently server-class ARM-based CPUs are also becoming available, both on-prem and in the cloud. In this paper we present our experience in running a sample CORSIKA simulation on both ARM and x86 CPUs available through Google Kubernetes Engine (GKE). We used the production binaries for the x86 instances, but had to build the binaries for ARM instances from source code, which turned out to be mostly painless. Our benchmarks show that ARM-based CPUs in GKE were not only the most cost-effective but were also the fastest in absolute terms in all the tested configurations. While the advantage is not drastic, about 20% in cost-effectiveness and less than 10% in absolute terms, it is still large enough to warrant an investment in ARM support for IceCube.

## 1 Introduction

The IceCube Neutrino Observatory [1] is the world's premier facility to detect neutrinos with energies above 1 TeV and an essential part of multi-messenger astrophysics. It instrumented a cubic kilometer of naturally occurring ice at the south pole, where it looks for light that is emitted when neutrinos interact the ice. IceCube needs to simulate the detector response to both background and signal to enable research to create algorithms to separate the signal from the background and study how the detector inherent features could affect the results. The background in IceCube is dominated by muons from cosmic ray air showers produced in the atmosphere above IceCube.

IceCube uses CORSIKA [2] to produce the necessary cosmic ray air shower simulations [3]. Millions of CPU core hours are needed to produce the necessary statistics needed for IceCube's needs. Historically, CORSIKA was compiled and run exclusively on x86-based CPUs, like Intel Xeon and AMD EPYC, but the source code itself makes no explicit assumptions about the CPU architecture; the x86-based CPUs were simply the most widely available and the most cost-effective solution available, both on-prem and in the cloud.

Recently, however, several cloud providers started offering ARM-based CPUs in their offering. This makes ARM CPUs easily available at scale to the science community, both as an option for occasional cloud-bursting and as an evaluation option before purchasing similar equipment on-prem.

We thus set up a Google Kubernetes Engine (GKE) instance and configured several node pools, each with its own instance type. We picked instances providing 4 vCPUs each, and then started Kubernetes pods that ran 4 parallel sequences of simulations, for which we collected the run times.

The details of the setup are available in Section 2 and the benchmark results are available in Section 3.

## 2 Using CORSIKA on GKE

We decided to use a Kubernetes-based setup, since it makes for an easily repeatable setup. IceCube software, and in particular CORSIKA, can run on any Linux-based system, so we used the GKE-provided base OS container images, in particular centos:centos7 for x86-based CPUs and almalinux:8.6 for ARM-based CPUs.

We created several node pools, each with its own machine type. The ARM-based machine type was t2a-standard-4, which is based on the Ampere Altra CPU. The closest x86-based machine type was t2d-standard-4, which is based on the AMD EPYC CPU. For complete comparison, we measured the runtimes also on the hyperthreading-enabled c2d-standard-4 and c2-standard-4 machine types. A summary of the machine types in the us-central1 region at the time of testing (August 2022) is provided in Table 1, while current detailed specs can be found on Google's web site [4].

Table 1. Summary of the tested GKE machine types.

| Machine type | CPU type | vCPUs | Physical cores | Core frequency (Ghz) | On-demand cost ($/h) |
|---|---|---|---|---|---|
| t2a-standard-4 | ARM Altra Q64-30 | 4 | 4 | 3.0 all core | 0.154 |
| t2d-standard-4 | x86_64 AMD EPYC 7B13 | 4 | 4 | 2.45 base 2.8 effective 3.5 max boost | 0.169 |
| c2d-standard-4 | x86_64 AMD EPYC 7B13 | 4 | 2 | 2.45 base 2.8 effective 3.5 max boost | 0.1816 |
| c2-standard-4 | x86_64 Intel® Xeon® Gold 6253CL | 4 | 2 | 3.1 base 3.8 all core 3.9 single core | 0.2088 |

We used the available x86-based binaries from the production CVMFS read-only shared filesystem, which we mounted inside the pods [5]. Since IceCube did not have any ARM-based binaries readily available, we built them in a dedicated build pod on GKE, using the same tools and procedures used for the x86-based builds, and then exported those binaries to the other ARM-based pods. Building the IceCube software on an ARM-based CPU was relatively painless; most of the software just compiled without any problems, but we did have to replace some of the x86-specific options that were hardcoded with their ARM equivalents. Note that the CORSIKA build did not require any tweaking.

We deployed one Kubernetes pod per provisioned node. Each pod was configured to mount the architecture-appropriate binaries, alongside a script that would run 5 identical simulations in sequence. Four instances of the benchmark script were launched concurrently in each pod, with the timing logs of each saved in a separate location. Once all the activity in a pod was completed, the logs were collected, and the pod deleted.

The ARM-based CPUs were still in preview in GKE, so we had to use a rapid-channel Kubernetes version. The runtime numbers were collected using version 1.23.8-gke.400.

## 3 Benchmark results

In this section we present the analysis of the collected timing logs. To minimize the effect of start-up and ramp-down activity, we keep only the central timing data points for each script, discarding the first and the last timing entry.

We independently collected timing data of three standard simulation configurations [6]. They cover low (0.6 – 30 TeV), medium (30 – 1k TeV) and high energy (1k – 1M TeV) cosmic ray energy simulations. All three exhibited approximately the same runtime ratio between the tested machine types, so due to limited space availability, we will present the details of only one here.

The min, max and mean runtime values of the low energy simulation are available in Table 2. As can be seen, simulations running on the ARM-based t2a-standard-4 machine type were the fastest and also exhibited very low variability. The x86-based t2d-standard-4 is not significantly slower, especially at the lower end, but it does exhibit significantly more variability. The hyperthread-enabled c2d-standard-4 and c2-standard-4 are instead significantly slower, which is not surprising, since they effectively provide half of the physical CPU cores.

**Table 2.** Summary run times for IceCube's low energy simulations.

| Margin | CPU type | Run time, in minutes | | |
|---|---|---|---|---|
| | | Mean | Min | Max |
| t2a-standard-4 | Altra ARM | 38.4 | 37.7 | 39.8 |
| t2d-standard-4 | AMD x86_64 | 41.5 | 38.2 | 45.6 |
| c2d-standard-4 | AMD x86_64 | 51.0 | 48.7 | 52.5 |
| c2-standard-4 | Intel x86_64 | 55.2 | 55.0 | 55.4 |

The ARM-based machine types were also more cost effective to rent than the x86-based ones in GKE. Table 3 provides the number of simulations we can perform on average in each, both per hour and per dollar. As can be clearly seen, the ARM-based machine types deliver significantly more simulation results for the same budget in GKE.

Table 3. Cost effectiveness for IceCube's low energy simulations.

| Margin | CPU type | Simulations/h | Simulations/$ |
|---|---|---|---|
| t2a-standard-4 | Altra ARM | 6.2 | 40 |
| t2d-standard-4 | AMD x86_64 | 5.8 | 34 |
| c2d-standard-4 | AMD x86_64 | 4.7 | 26 |
| c2-standard-4 | Intel x86_64 | 4.3 | 21 |

## 4 Summary and conclusions

ARM-based CPUs are becoming widely available in production server environments. IceCube has substantial CPU compute needs but has used exclusively x86-based CPUs in the recent past. We thus decided to evaluate the feasibility, performance, and cost-effectiveness of adding ARM-based CPUs to its resource mix.

At the time of testing, the easiest way to access a set of ARM CPUs was by provisioning them from a commercial cloud. We picked Google Kubernetes Engine as our testbed, both for its ease of use and the availability of both ARM-based and x86-based CPUs. The well-defined pricing structure also made for an excellent environment in which to compare cost-effectiveness of the various CPUs.

Our benchmarks show that ARM-based CPUs were not only the most cost-effective but were also the fastest in absolute terms. While the advantage is not drastic, about 20% in cost-effectiveness and less than 10% in absolute terms, it is still large enough to warrant an investment in ARM support for IceCube.

Especially since building IceCube's software for the ARM-based CPU was trivial. Most software components just built out of the box. We did find a few places where the build script had x86-specific options, but those were easy to replace with their ARM equivalents. We plan to integrate the appropriate smarter logic in the official build scripts in the near future.

In summary, we were pleasantly surprised by the performance and value of the tested ARM-based CPUs. The test result clearly show that x86-based CPUs are not the only worthy option anymore and that we should start supporting ARM-based CPUs as first-class citizens in server-class compute.

## Acknowledgments

This work was partially funded by the U.S. National Science Foundation (NSF) under grants OAC-1826967, OAC-1541349, OPP-2042807, CNS-1925001 and OAC-2103963. All Google Kubernetes Engine costs have been covered by Google-issued credits.

## References

1. M. G. Aartsen et al., *The IceProd framework: Distributed data processing for the IceCube neutrino observatory*, Journal of Parallel and Distributed Computing 75 198-211 (2015). ISSN 0743-7315. https://doi.org/10.1016/j.jpdc.2014.08.001
2. D. Heck et al., *CORSIKA: A Monte Carlo code to simulate extensive air showers,* FZKA-6019 (1998) https://doi.org/10.5445/IR/270043064
3. D. Soldin et al., *Atmospheric Muons Measured with IceCube*, EPJ Web Conf., 208 (2019) 08007 https://doi.org/10.1051/epjconf/201920808007
4. Google Cloud Compute Engine, *About machine families*, online, accessed Aug 2022. https://cloud.google.com/compute/docs/machine-types
5. I. Sfiligoi, D. Schultz, B. Riedel, F. Wuerthwein, S. Barnet, and V. Brik, *Demonstrating a Pre-Exascale, Cost-Effective Multi-Cloud Environment for Scientific Computing: Producing a fp32 ExaFLOP hour worth of IceCube simulation data in a single workday,* In Practice and Experience in Advanced Research Computing (PEARC '20). Association for Computing Machinery, New York, NY, USA, 85-90. (2020) https://doi.org/10.1145/3311790.3396625
6. GitHub, Benchmark scripts, online, accessed Aug 2022. https://github.com/sfiligoi/tnrp-net-tests/tree/master/gke22-icecube-arm/scripts